\documentclass[amsmath,tightenlines,superscriptaddress,twocolumn,amssymb,prb,aps]{revtex4}

\usepackage{graphicx}
\usepackage{dcolumn}
\usepackage{pdfsync}
\usepackage{bm}
\usepackage{array}
\usepackage{hyperref}
\usepackage{multirow}

\begin{document}
\title{Novel high pressure structures and superconductivity of niobium disulfide}
\author{Zhong-Li Liu}
\email{zl.liu@163.com}

\affiliation{College of Physics and Electric Information, Luoyang Normal University, Luoyang 471022, China}
\affiliation{Laboratory for Shock Wave and Detonation Physics Research, Institute of Fluid Physics, P.O. Box 919-102, 621900 Mianyang, Sichuan, China}

\author{Ling-Cang Cai}

\affiliation{Laboratory for Shock Wave and Detonation Physics Research, Institute of Fluid Physics, P.O. Box 919-102, 621900 Mianyang, Sichuan, China}

\author{Xiu-Lu Zhang}

\affiliation{Laboratory for Extreme Conditions Matter Properties, Southwest University of Science and Technology, 621010 Mianyang, Sichuan, China}

\date{\today}

\begin{abstract}
We have investigated the pressure-induced phase transition and superconducting properties of niobium disulfide (NbS$_2$) based on the density functional theory. The structures of NbS$_2$ at pressures from 0 to 200 GPa were predicted using the multi-algorithm collaborative (MAC) structure prediction technique. The previously known 1$T$-, 2$H$-, and 3$R$-NbS$_2$ were successfully reproduced. In addition, many metastable structures which are potential to be synthesized were also discovered. Based on the enthalpy calculations, we found that at 26 GPa NbS$_2$ transits from the double-hexagonal (2$H$) structure to the tetragonal $I4/mmm$ structure with a 10.6\% volume reduction. The calculated elastic constants and phonon dispersion curves of $I4/mmm$-NbS$_2$ confirm its mechanical and dynamical stability at high pressure. More interestingly, the coordination number of Nb in $I4/mmm$ structure is eight which is larger than that in the traditional metal dichalcogenides, indicating a new type of bondings of Nb and S atoms. In the new Nb-S bondings, one Nb atom and neighboring eight S atoms form a [NbS$_8$] hexahedron unit. Furthermore, $I4/mmm$-NbS$_2$ exhibits a higher superconducting critical temperature than 2$H$-NbS$_2$, as is resulted from the stronger electron-phonon coupling coefficients. \end{abstract}

\pacs{}
  
\maketitle

\section{Introduction}
Transition metal dichalcogenides (TMDs) MX$_2$ (M = Nb, Ta, Mo, W, X = S, Se, Te) have  intriguing properties, ranging from insulator to metal and superconductor, and thus always attract extensive interests of experimentalists and theorists. Thanks to their in-plane covalent bondings and weak interlayer van der Waals interactions, they could be easily exfoliated down to a monolayer which shows very exotic properties. For example, bulk MoS$_2$ is an indirect-band-gap semiconductor,~\cite{Wei2010,Ding2012} while the monolayer MoS$_2$ is a direct-band-gap semiconductor.~\cite{Lebegue2009} Consequently, the TMDs have shown exciting prospects for a variety of applications, such as catalysts and lubricants in the petroleum industry~\cite{Raybaud1997}, promising applications in nanoelectronics and optoelectronics,~\cite{Wang2012} and energy storage applications.~\cite{Chhowalla2012}

The typical representative of TMDs is 2$H$-NbSe$_2$, showing a large charge density wave (CDW) (at 33 K) that coexists with superconductivity ($T_c$ = 7.2 K).~\cite{Moncton1977} Niobium disulfide (NbS$_2$) also belongs to the family of TMD compounds.  But the CDW order appeared in 2$H$-NbSe$_2$ is absent in 2$H$-NbS$_2$,~\cite{Guillamon2008,Kacmarcik2010,Leroux2012} and its occurrence is suppressed by the large anharmonic effects.~\cite{Leroux2012} However, it also shows superconductivity at a similar transition temperature of $T_c$ = 6 K.~\cite{Wilson1975,Hamaue1986,Tissen2013a} More interestingly, the T$_c$ of 2$H$-NbS$_2$ increases smoothly from 6 K at zero pressure to $\sim$8.9 K at 20 GPa,~\cite{Tissen2013a} also similar to the behavior of $T_c$ in 2$H$-NbSe$_2$ which increases to $\sim$ 8.5 K at 10 GPa.~\cite{Tissen2013a} The upper critical field of 2$H$-NbS$_2$ has an initial decrease as pressure increases, contrary to the increase of $T_c$, but above 8.7 GPa it increases again with pressure.~\cite{Tissen2013a} 

NbS$_2$ is a two-gap superconductor, similar to NbSe$_2$. The heat capacity of a 2$H$-NbS$_2$ has been measured down to 0.6 K and in magnetic fields up to 14 T by Ka$\mathrm{\check{c}}$mar$\mathrm{\check{c}}$ik \textit{et al}.~\cite{Kacmarcik2010} The temperature dependence of the electronic specific heat can be attributed to either the existence of a strongly anisotropic single-energy gap or a two-gap scenario with the large gap about twice bigger than the small one. The field dependence of the Sommerfeld coefficient induces a magnetic field dependence of the superconducting anisotropy.~\cite{Kacmarcik2010} The two-gap scenario conclusions are supported by the absence of in-plane gap anisotropy in recent STM imaging of the vortex lattice in NbS$_2$.~\cite{Guillamon2008}

2$H$-NbS$_2$ has a layered structure and therefore has large anisotropic electrical, optical, and magnetic properties. It has been applied as catalyst for the purification of petroleum,~\cite{Geantet1996} cathode materials in secondary batteries,~\cite{Kumagai1991} humidity sensors,~\cite{Divigalpitiya1990,Ge2013} and so on. In experiment, presently the one-layer trigonal 1$T$-NbS$_2$,~\cite{Carmalt2004} two-layer hexagonal 2$H$-NbS$_2$,~\cite{Jellinek1960} and three-layer rhombohedral 3$R$-NbS$_2$~\cite{Clark1976,Onari1979} heve been synthesized. Large-scale synthesis of 3$R$-NbS$_2$ nanosheets has also been recently realized.~\cite{Ge2013} Different low dimensional structures of NbS$_2$ have different physical and chemical properties. Low dimensional materials depend on and can be exfoliated from bulk materials. It is necessary to uncover as many crystal structures of NbS$_2$ as possible. From some new crystals, it is expected to exfoliate some new-functional low dimensional materials.

It is known that pressure is able to modulate the properties of materials through changing their crystal structures. Furthermore, the structures of NbS$_2$ under high pressure are fundamental to understand its superconductive properties. The mechanism of pressure-induced superconductivity and the superconducting temperature in NbS$_2$ above 20 GPa still remain unknown to us. This motivates us to investigate the superconductivity of NbS$_2$ at higher pressures. In this work, we first predicted the high-pressure structures of NbS$_2$ and determined its phase transition sequence using the multi-algorithm collaborative (MAC) crystal structure prediction technique combined with the density functional theory (DFT). Then we calculated the superconducting critical temperature through electron-phonon coupling calculations.

The paper is organized as follows. Section~\ref{compdet} contains the computational details. The results and discussion are presented in Sec.~\ref{results}. Conclusions follow in Sec.~\ref{concl}.

\section{Computational details}
\label{compdet}

In order to determine the high-pressure structures of NbS$_2$, we searched its low-energy structures from 0 to 200 GPa using our developed MAC crystal structure prediction technique.~\cite{muse} The multi algorithms including the evolutionary, the simulated annealing, and the basin hopping algorithms are combined to collaboratively search the global energy minima of materials with the fixed stoichiometry. The MAC algorithm and all the relevant techniques are incorporated in the {\sc Muse} code.~\cite{muse} The results were also carefully cross checked and confirmed by the {\sc Calypso} code,~\cite{Wangyc2010,Wangyc2012,Lv2012} which is based on the particle swarm optimization algorithm.

The \textit{ab initio} optimizations for every structure generated by the {\sc Muse} code were performed with {\sc vasp} package.~\cite{Kresse94,Kresse96} We tested the local density approximation (LDA) and the generalized gradient approximation (GGA) parametrized by Perdew, Burke and Ernzerhof (PBE)~\cite{Perdew1996} for exchange– correlation energy. The two approximations give the similar structures order in structure prediction. While the LDA calculated lattice constants are better than the GGA for NbS$_2$. So in the static calculations, we adopted the LDA exchange-correlation functional. The electron-ion interactions are described by the projector augmented wave (PAW) scheme.~\cite{Blochl1994,Kresse99} The pseudopotentials for Nb and S have the valence electrons' configurations of $\mathrm{4p^65s^14d^4}$ and $\mathrm{3s^23d^4}$, respectively. To achieve good convergences the kinetic energy cutoff and the \textit{k}-point grids spacing were chosen to be 500 eV and 0.02 $\mathrm{\AA}^{-1}$, respectively. The accuracies of the target pressure and the energy convergence for all optimizations are better than 0.1 GPa and $10^{-5}$ eV, respectively.  The searched systems contained 6, 9, 12, 15 and 18 atoms in the unit cell.

\section{Results and discussion}
\label{results}
\subsection{Structure prediction for NbS$_2$ at high pressures}
In our MAC structure searches,~\cite{muse} the structures were generated with symmetry constraints in the first generation and optimized with {\sc vasp} package at fixed pressures. The pressures applied to crystal structures in optimizations go from 0 to 200 GPa with the interval of 20 GPa. At each fixed pressure, the enthalpies of these structures were calculated and compared to find the proper path towards the lowest-enthalpy structure. Results show that the previously known 2$H$-NbS$_2$ has the lowest enthalpy at lower pressures (below 20 GPa) and the $I4/mmm$ structure has the lowest enthalpy at higher pressures (above 20 GPa). The previously known 1$T$- and 3$R$-NbS$_2$ structures were also easily reproduced. The large-scale 3$R$-NbS$_2$ nanosheets are synthesized very recently.\cite{Ge2013} More interestingly, we found a new two-layer hexagonal structure whose energy is very close to 3$R$-NbS$_2$ at ambient pressure. We refer to it as 2$H$$'$-NbS$_2$. According to energy criterion, it is potential to be synthesized in experiment. Meanwhile, many other structures were found to be energetically competitive, including $P3m1$, $P6_422$, $P6_222$ structures, and so on. Among these structures, the trigonal $P3m1$ structure has lower energy with respective to the known 3$R$-NbS$_2$ in the whole pressure range of interest. So, it is also expected to be synthesized in experiment.

\begin{table}
\caption{The comparison of the calculated lattice constants $a$, $c$, and $c/a$ of different NbS$_2$ structures with the corresponding experimental values.}
\label{tab:lattcont}

\begin{tabular}{ccccccc}

\hline\hline
Structure & Method & $a (\mathrm{\AA})$ & $c (\mathrm{\AA})$ & $c/a$ & $P$ (GPa) & Reference \\
\hline
1$T$-NbS$_2$ & VASP-LDA & 3.253 & 5.341 & 1.642 &0& This work \\ 
& Experiment & 3.420 & 5.938& 1.736 &0& \onlinecite{Carmalt2004}\\
2$H$-NbS$_2$ & VASP-LDA & 3.287 & 11.421 & 3.475  &0& This work \\ 
& Experiment & 3.418 & 11.860 & 3.470 & 0& \onlinecite{Carmalt2004}\\
& Experiment & 3.310 & 11.890 & 3.592 &0& \onlinecite{Jellinek1960}\\
& Experiment & 3.330 & 11.950 & 3.589 &0& \onlinecite{Leroux2012}\\
3$R$-NbS$_2$ & VASP-LDA & 3.286 & 17.577 & 5.349 &0& This work \\ 
& Experiment & 3.335 & 17.834 & 5.336 & 0&\onlinecite{Ge2013}\\
\hline\hline
\end{tabular}
\end{table}

\begin{figure}[h]
\begin{center}
\includegraphics*[width=9.0cm]{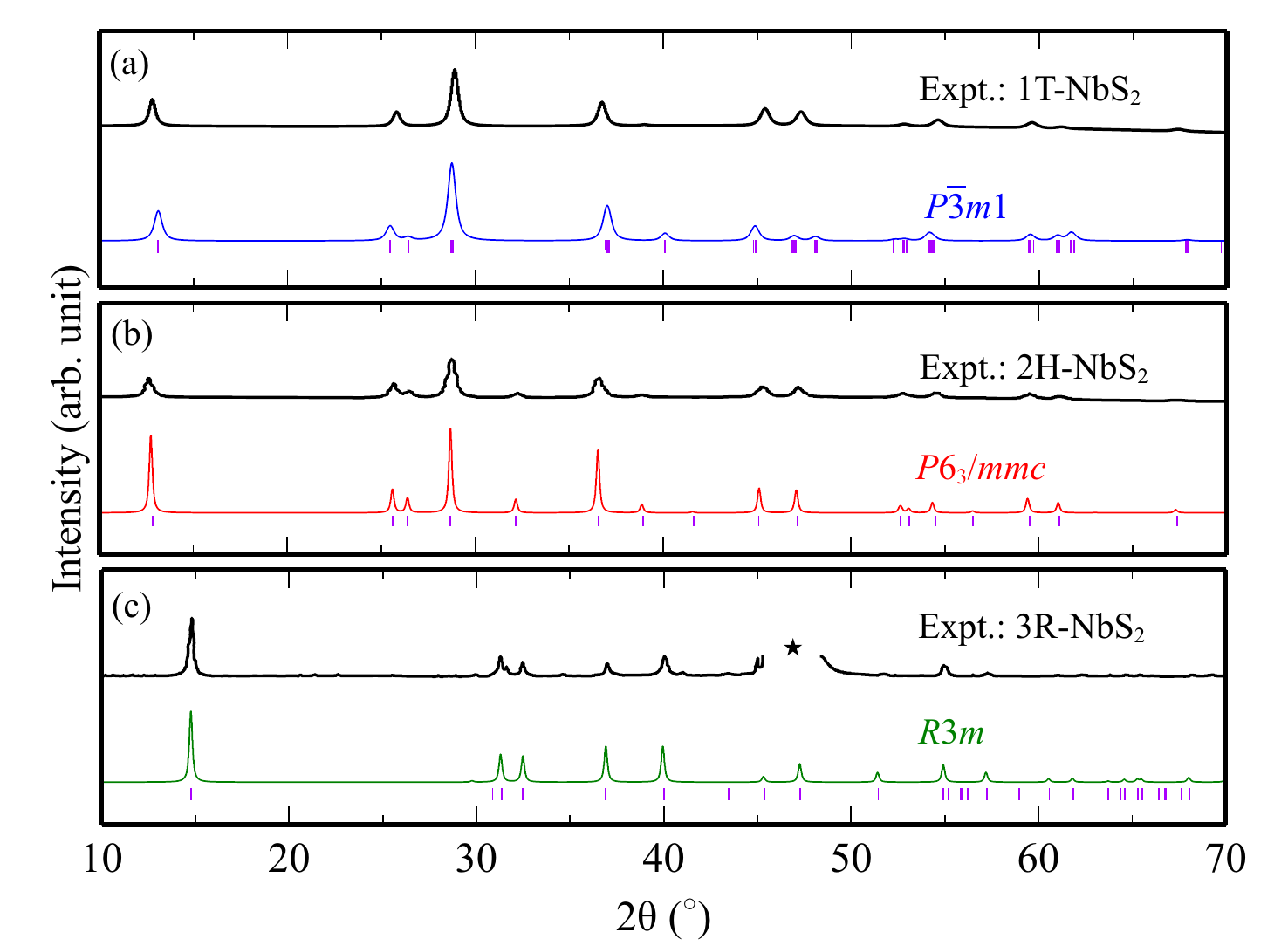}
\caption{(color online). Simulated XRD patterns of 2$H$-NbS$_2$, 3$R$-NbS$_2$, and 1$T$-NbS$_2$, in comparison with the corresponding experimental results (1$T$: \onlinecite{Carmalt2004}, 2$H$: \onlinecite{Carmalt2004}, 3$R$: \onlinecite{Ge2013}).}
\label{fig:xrd}
\end{center}
\end{figure}

The calculated lattice constants of 1$T$-, 2$H$-, and 3$R$-NbS$_2$ are listed in Table.~\ref{tab:lattcont}, in comparison with experimental values.~\cite{Carmalt2004,Jellinek1960,Leroux2012,Ge2013} We note that the lattice constants $a$ and $c$ of the three structures are all slightly underestimated in our LDA calculations. But the calculated $c/a$ values are all in good agreement with experiments. To further examine the three structures, we also simulated their X-ray diffraction (XRD) patterns and compare them with experimental data. The calculated XRD patterns of the three structures are all in excellent agreement with corresponding experiments (Fig.~\ref{fig:xrd}), indicating that each structure is identical to the known one. 

\begin{figure}[h]
\begin{center}
\includegraphics*[width=9.0cm]{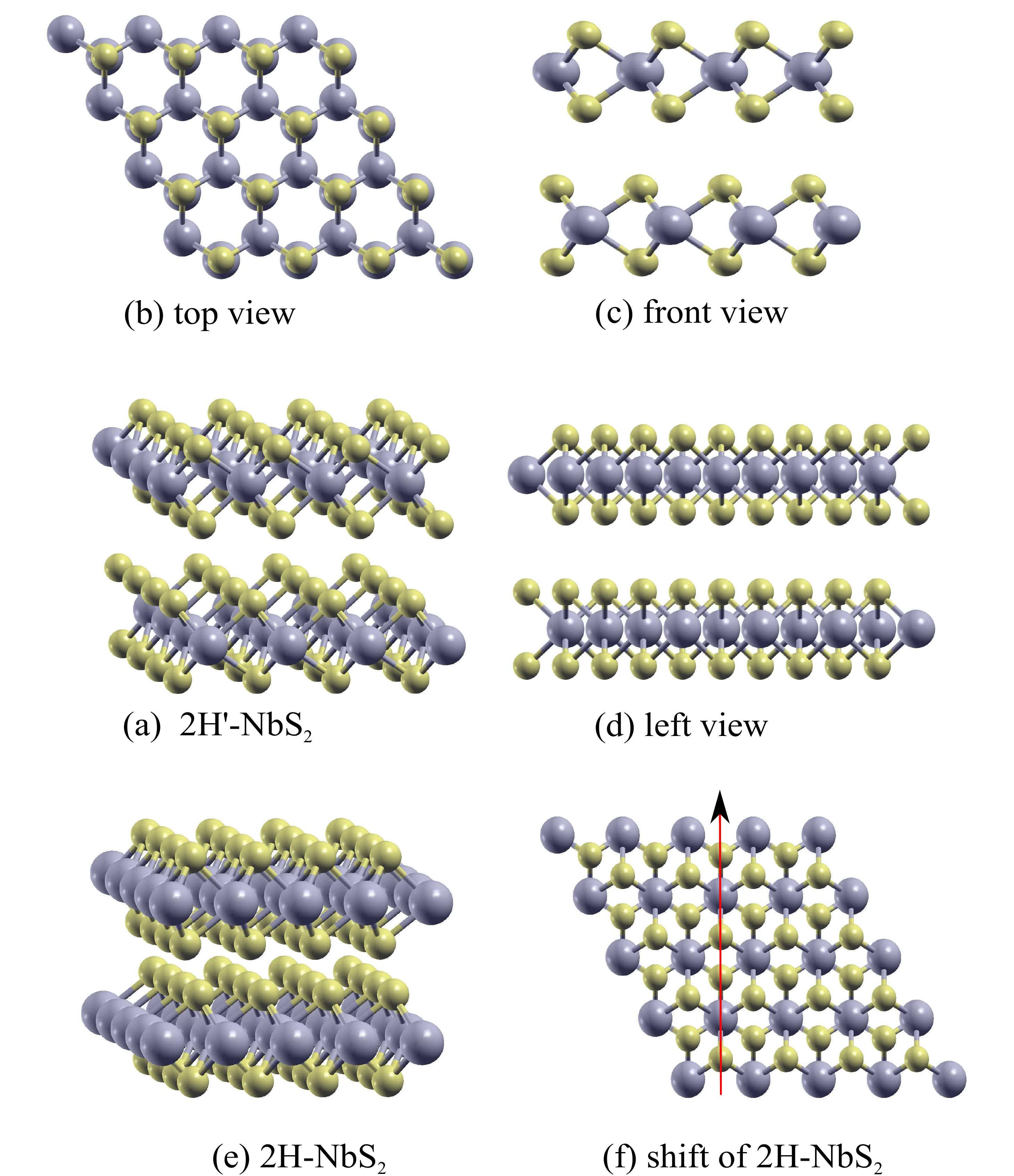}
\caption{(color online). The 2$H$-NbS$_2$ and 2$H'$-NbS$_2$ crystal structures. (a) The structure of 2$H'$-NbS$_2$. (b) Top view of 2$H'$-NbS$_2$. (c) Front view of 2$H'$-NbS$_2$. (d) Left view of 2$H'$-NbS$_2$. (e) The structure of 2$H$-NbS$_2$. (f) The shift direction of one layer of 2$H$-NbS$_2$ to form 2$H'$-NbS$_2$ (top view).}
\label{fig:2Hp}
\end{center}
\end{figure}

The new 2$H'$-NbS$_2$ crystal has the 2$H$-MoS$_2$ structure and can be formed by shifting one layer of atoms in 2$H$-NbS$_2$. The shifting distance is 0.577 lattice constant $a$ along typical direction. 2$H'$-NbS$_2$ has six atoms in primitive cell with the lattice constants of 3.28, 3.28 and 11.65 $\mathrm{\AA}$ at ambient pressure. The Nb and S atoms are at Wyckoff's 2c positions (1/3, 2/3, 1/4) and 4f positions (1/3, 2/3, 0.62), respectively. We show the 2$H'$-NbS$_2$ structure and the shifting direction in Fig.~\ref{fig:2Hp}. The shifting direction is parallel to the layer plane (Fig.~\ref{fig:2Hp} f). That is to say that the structures of the two layers are the same. The unique difference between 2$H'$- and 2$H$-NbS$_2$ is the relative positions of the two layers. The coordination numbers of Nb atoms in both 2$H'$- and 2$H$-NbS$_2$ are six. One Nb atom and the neighboring six S atoms form a [NbS$_6$] trigonal prismoid. Accordingly, the coordination numbers of S atoms in both 2$H'$- and 2$H$-NbS$_2$ are three.

\begin{figure}[h]
\begin{center}
\includegraphics*[width=9.0cm]{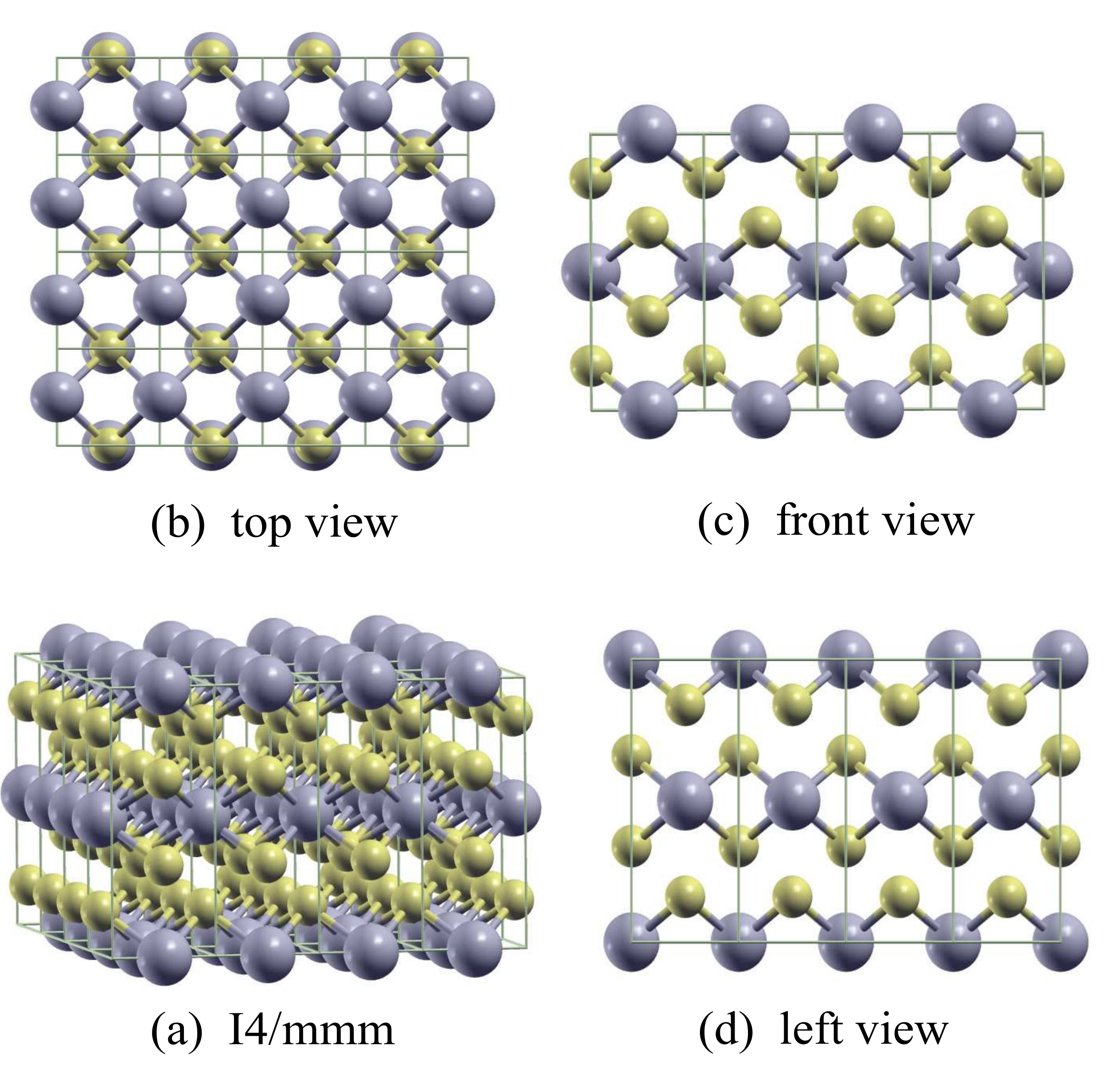}
\caption{(color online). Predicted $I4/mmm$ crystal structure. (a) The structure of $I4/mmm$-NbS$_2$. (b) Top view. (c) Front view. (d) Left view.}
\label{fig:139}
\end{center}
\end{figure}

The new $I4/mmm$ structure is plotted in Fig.~\ref{fig:139}. It has six atoms in conventional unit cell (three in primitive cell) with the lattice constants of 3.15, 3.15 and 7.91 $\mathrm{\AA}$ at ∼26 GPa. The Nb and S atoms are at Wyckoff's 2a positions (0.0, 0.0, 0.0) and 4e positions (0.0, 0.0, 0.34), respectively.  More interestingly, the coordination number of Nb in $I4/mmm$ is eight. In this new type of bondings, one Nb atom and neighboring eight S atoms form a [NbS$_8$] hexahedron. The coordination number of S is four. To our knowledge, this type of covalent bondings has not been reported in TMD crystals. In general, in TMDs the metal atom has traditional six nearest neighbors.~\cite{Chhowalla2012} This new type of eight nearest neighbors in $I4/mmm$-NbS$_2$ implies new potential chemical and physical properties, especially in two-dimensional crystals.

\subsection{Phase transition and structural stability of NbS$_2$}

\begin{figure}[h]
\begin{center}
\includegraphics*[width=9.0cm]{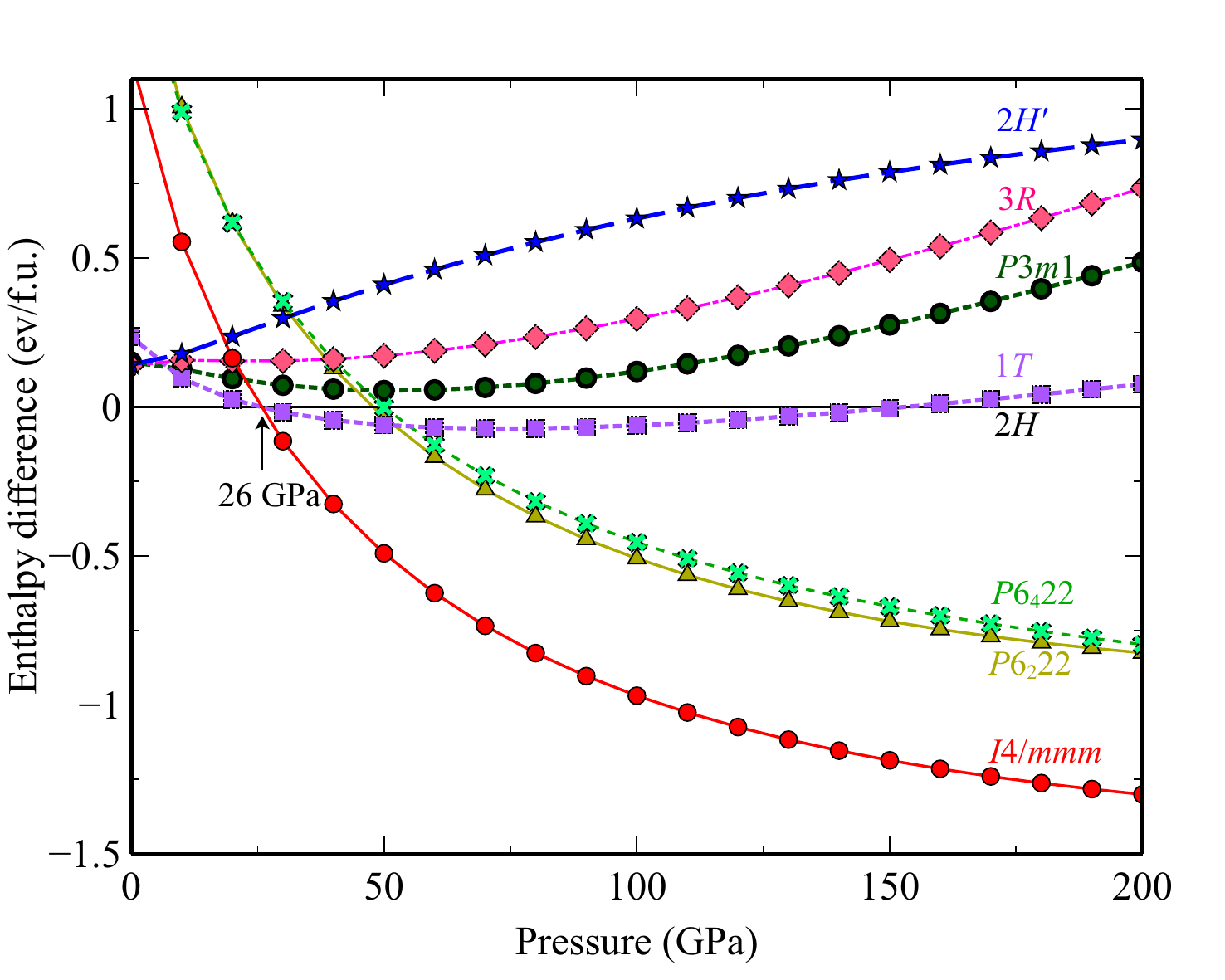}
\caption{(color online). Enthalpy differences of predicted structures relative to 2$H$-NbS$_2$ structure under high pressure.}
\label{fig:enth}
\end{center}
\end{figure}

\begin{figure}[h]
\begin{center}
\includegraphics*[width=9.0cm]{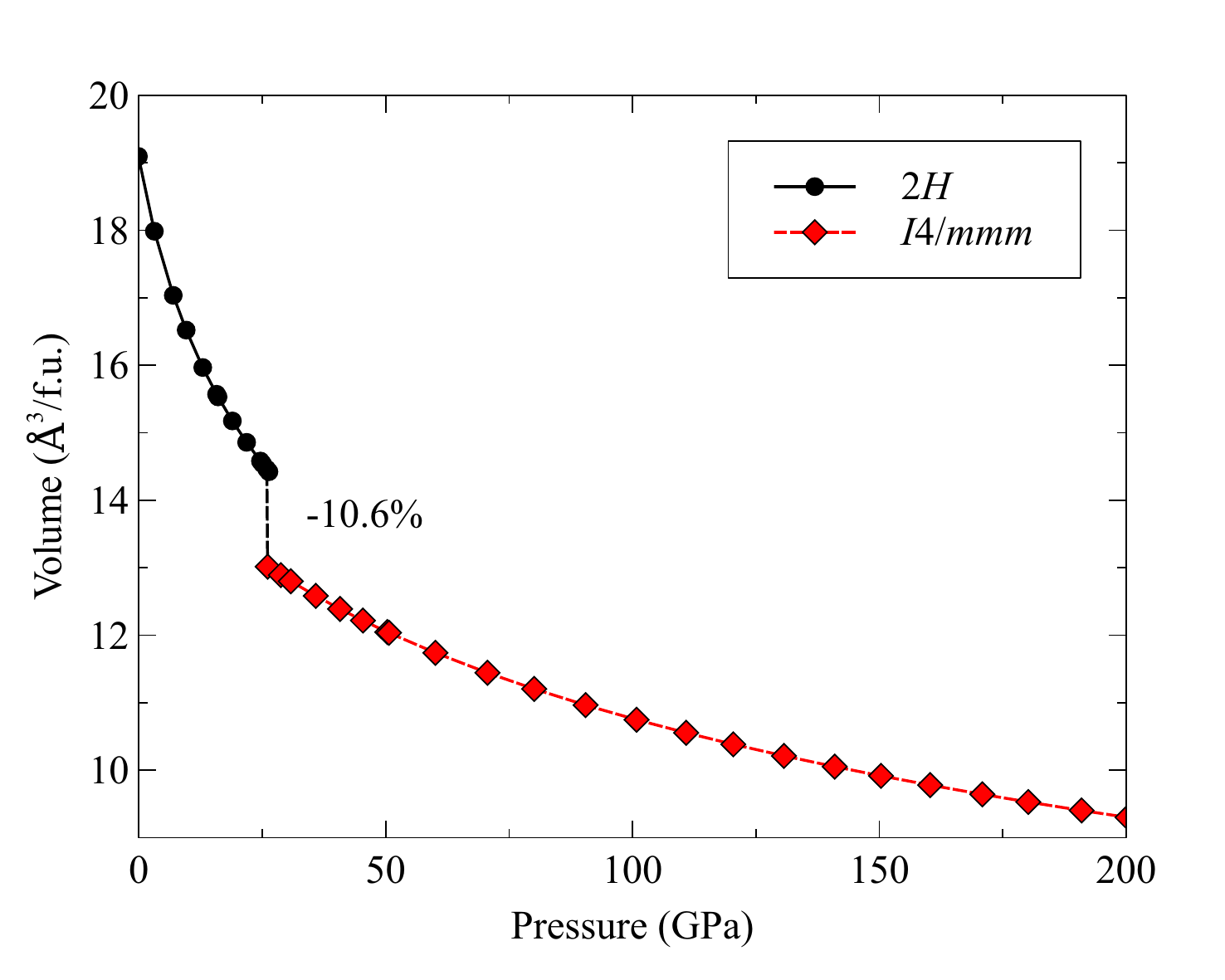}
\caption{(color online). The equation of states of NbS$_2$. The vertical dash curve indicates the volume reduction of NbS$_2$ at the phase transition point, 26 GPa.}
\label{fig:pv}
\end{center}
\end{figure}

In order to obtain the phase-transition sequence of NbS$_2$ under compression, we calculated the energies for its different phases at 0 K and pressures from 0 to 200 GPa. The enthalpies vs pressure data of different structures with respective to 2$H$-NbS$_2$ are plotted in Fig.~\ref{fig:enth}, from which we note at 0 K the previously known hexagonal 2$H$-NbS$_2$ is stable up to 26 GPa. Above 26 GPa, NbS$_2$ transits to the tetragonal $I4/mmm$ structure which remains stable up to a very high pressure, 200 GPa, the upper limit of our interest. Upon compression,  NbS$_2$ exhibits a volume reduction of 10.6\% at 26 GPa (Fig.~\ref{fig:pv}). This volume reduction directly results in the decrease of interlayer distance and the aggregation of S atoms around Nb atoms. Although at ambient conditions, 1$T$- and 3$R$-NbS$_2$ have relatively higher energies than 2$H$-NbS$_2$, they have been synthesized successfully. The energies of 2$H'$- and $P3m1$-NbS$_2$ are close to that of 3$R$-NbS$_2$, so we believe they are both potential to be synthesized in experiment. After all, the trigonal $P3m1$ structure has lower energy than the known 3$R$-NbS$_2$ in the whole pressure range. The energies of $P6_222$ and $P6_422$ structures are both much higher than that of $I4/mmm$ structure. So the Gibbs free energy barrier is to high for NbS$_2$ to overcome.


\begin{figure}[ht]
\begin{center}
\includegraphics*[width=9.0cm]{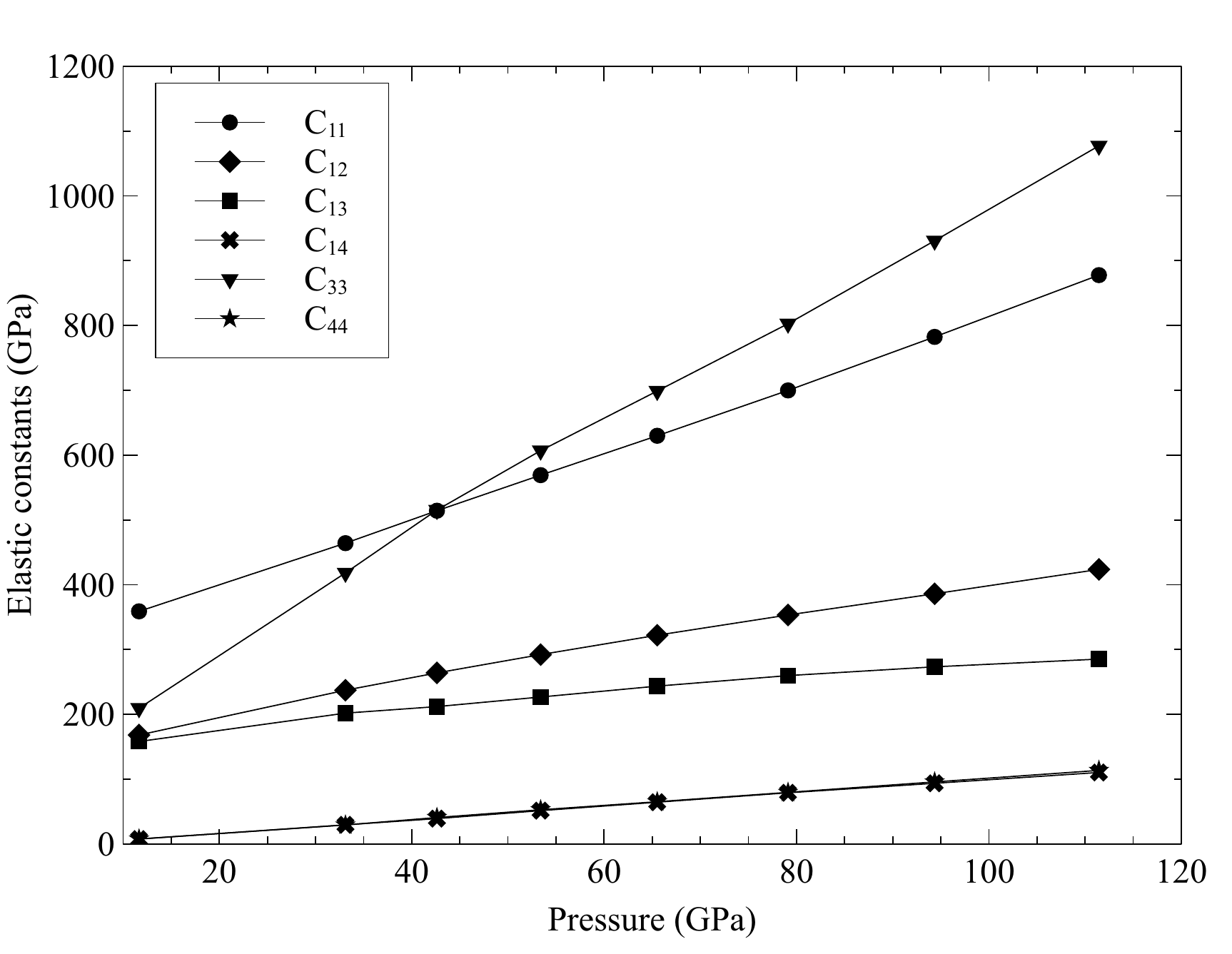}
\caption{The high-pressure elastic constants of $I4/mmm$-NbS$_2$.}
\label{fig:ec139}
\end{center}
\end{figure}

The mechanical stability of 2$H$- and $I4/mmm$-NbS$_2$ are confirmed by their elastic constants (shown in Table~\ref{tab:ec} and Fig.\ref{fig:ec139}) according to the elastic criteria of the hexagonal systems,~\cite{Wu2007}
\begin{equation}
C_{11}>|C_{12}|, \quad (C_{11}+2C_{12})C_{33}>2C_{13}^2, \quad C_{44}>0,
\end{equation}
and tetragonal systems,
\begin{equation}
\begin{aligned}
C_{11}>0, \quad C_{33}>0, \quad C_{44}>0, \quad C_{66}>0, \\
C_{11} > C_{12}, \quad C_{11}+C_{33}>2C_{13},\\
2(C_{11} + C_{12})+C_{33}+4C_{13}>0,
\end{aligned}
\end{equation} 
respectively. The increasing of the elastic constants of $I4/mmm$-NbS$_2$ with pressure reflect its enhanced stability as pressure increases (Fig.\ref{fig:ec139}). The new 2$H'$ structure is also mechanically stable at ambient and high pressure according to the elastic criteria of the hexagonal crystals.~\cite{Wu2007} While, the $P3m1$ structure is only stable at low pressures. It becomes mechanically unstable as pressure increases because of the appearance of negative shear modulus $C_{14}$. It is also worthy to note that the shear modulus $C_{44}$ of 2$H$-NbS$_2$ increases with pressure, but the $C_{44}$ values of 2$H'$-NbS$_2$ remain small as pressure increases. This implies 2$H$-NbS$_2$ is more stable than 2$H'$-NbS$_2$. So, it is easier to synthesize 2$H$-NbS$_2$ in experiment other than 2$H'$-NbS$_2$.

\begin{table}
\caption{The elastic constants of different NbS$_2$ structures under high pressure. The pressure ($P$) and elastic constants are all in GPa.}
\label{tab:ec}

\begin{tabular}{cccccccc}

\hline\hline
Structure & $P$ & $C_{11}$  & $C_{12}$ & $C_{13}$ & $C_{33}$& $C_{44}$ & $C_{14}$\\
\hline
2$H$&0.00   &174.16 &77.34  &9.70   &58.80  &65.26  & -\\
          &11.90  &247.87 &94.07  &48.92  &232.36 &99.38  & -\\
          &21.69  &314.77 &127.03 &70.09  &324.01 &139.63 & -\\
2$H'$&0.00 &163.26 & 83.75 & 15.49 & 52.47 & 2.35 & -\\
& 19.93 &267.04 & 121.63& 82.27& 314.33& 10.43  & -\\
& 47.17 &336.47 &186.91& 134.25&593.40 & 15.98  & -\\
$P3m1$ & 0.00 & 186.36 & 68.43 & 20.35 & 103.46 & 15.88 & 1.07 \\
& 13.73 & 262.74 & 81.66 & 57.44 & 272.51 & 64.46 & -10.50\\
& 23.10 & 324.97 & 104.56 & 72.18 & 369.49 & 94.47 & -20.06\\
\hline\hline
\end{tabular}
\end{table}

To further check the dynamical stability of the new structures, 2$H'$- and $I4/mmm$-NbS$_2$, we determined their vibrational frequencies using density functional perturbation theory (DFPT),~\cite{Baroni1987,Baroni2001} as implemented in the QUANTUM-ESPRESSO package.~\cite{pwscfcite} For the exchange-correlation functional we used the Perdew Zunger local- density approximation (LDA)~\cite{Perdew1981} and ultrasoft pseudopotential.~\cite{Vanderbilt1990} We applied the Vanderbilt ultrasoft pseudopotentials for Nb and S with the valence electrons configurations 4s$^2$4p$^6$4d$^2$5s$^2$ and 3$s^2$3p$^4$, respectively. The ultrasoft pseudopotentials were generated with a scalar-relativistic calculation.

\begin{figure}[ht]
\begin{center}
\includegraphics*[width=9.0cm]{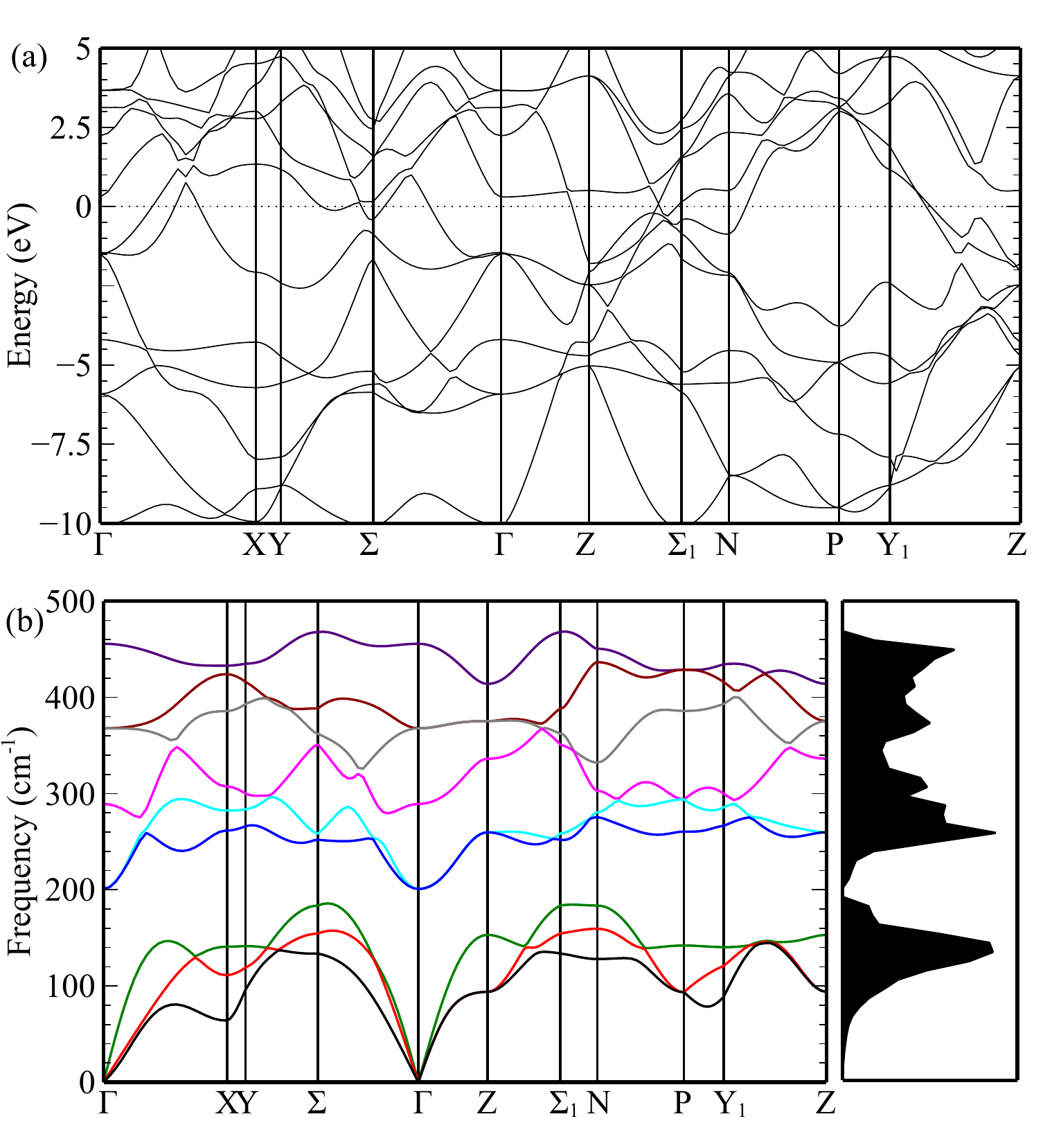}
\caption{(color online). The band structure and phonon dispersion curve of $I4/mmm$-NbS$_2$ at 60 GPa.  (a) the band structure, (b) the phonon dispersion curve (left) and phonon density of states (right).}
\label{fig:ph139}
\end{center}
\end{figure}

\begin{figure}[ht]
\begin{center}
\includegraphics*[width=9.0cm]{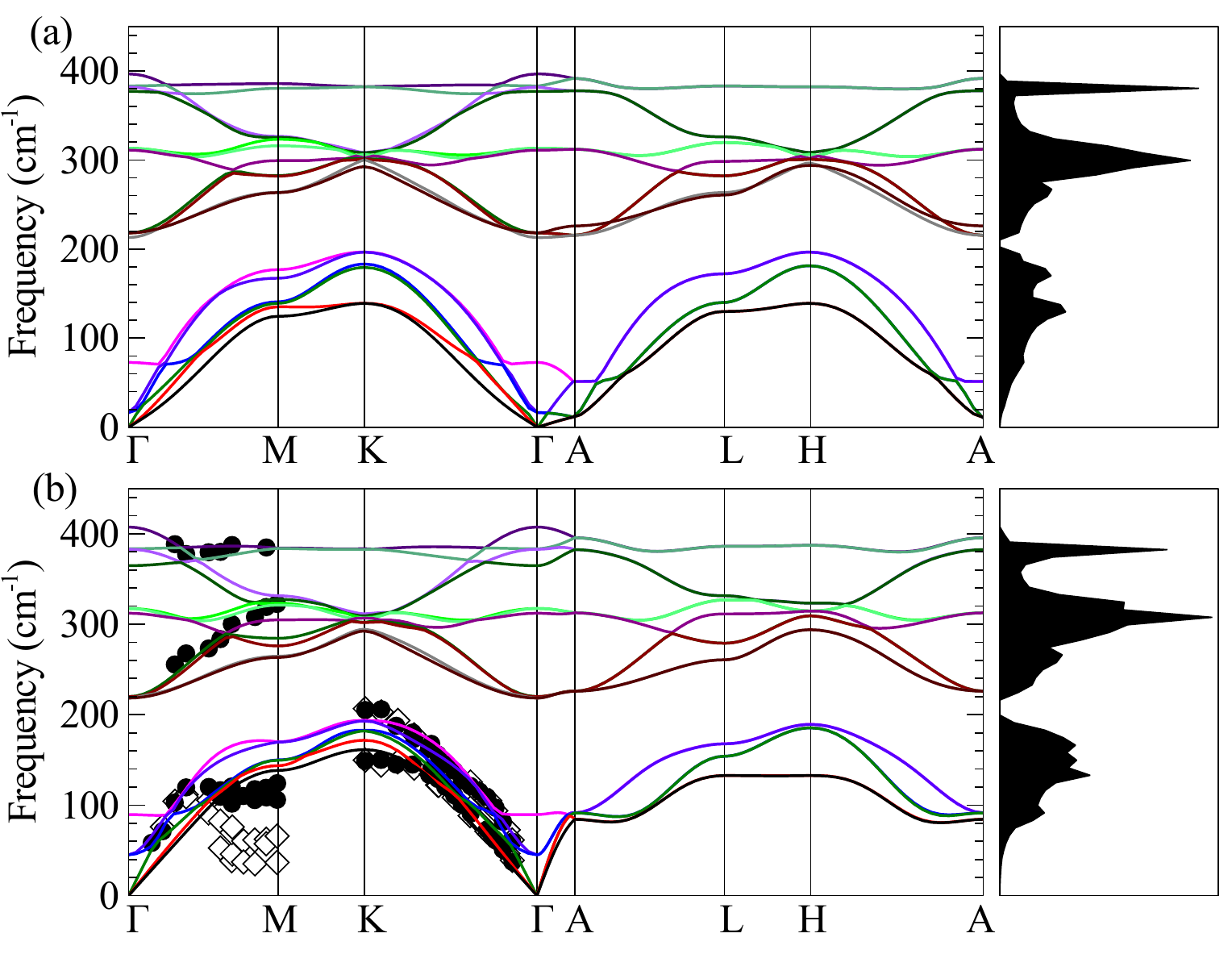}
\caption{(color online). The phonon dispersion curve of 2$H'$- (a) and 2$H$-NbS$_2$ (b) at 0 GPa (lines). The experimental data [Ref.~\onlinecite{Leroux2012}] of 2$H$-NbS$_2$ at 2 K (open diamonds) and 300 K (filled circles) are also plotted for comparison.}
\label{fig:ph194}
\end{center}
\end{figure}

\begin{figure}[!top]
\begin{center}
\includegraphics*[width=9.0cm]{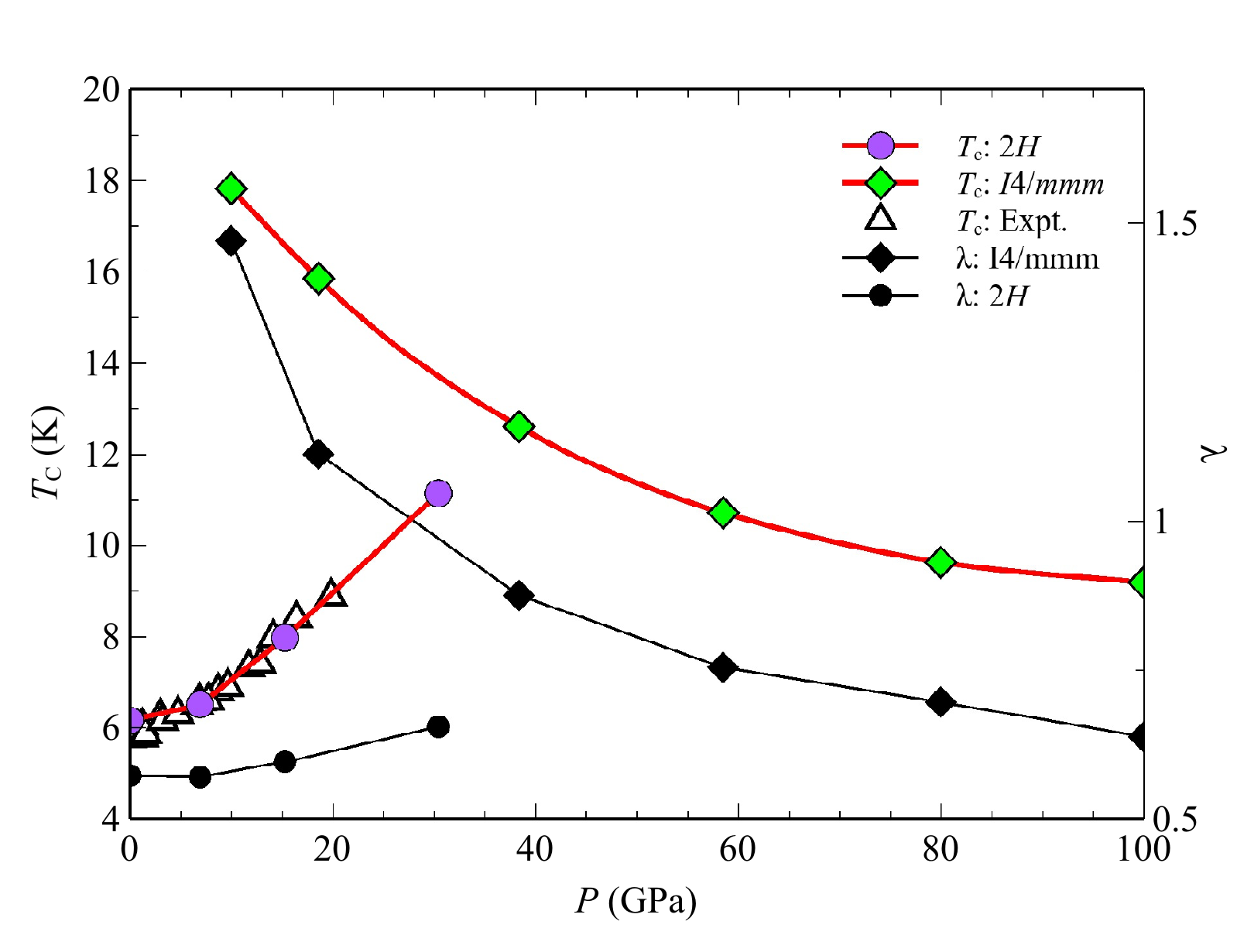}
\caption{(color online). The comparison of calculated superconducting critical temperatures with experimental results ~\onlinecite{Tissen2013}. The calculated electron-phonon coupling coefficients $\lambda$ are also plotted.}
\label{fig:tc}
\end{center}
\end{figure}

We careful tested on {\bf k} and {\bf q} grids, the kinetic energy cutoff, and other technical parameters to ensure good convergence of phonon frequencies. The kinetic energy cutoff, the energy  cutoff for the electron density, and the {\bf k} grids were chosen to be 40 Ryd., 400 Ryd., and   16$\times$16$\times$16 Monkhorst-Pack (MP)~\cite{Monkhorst1976} meshes in both total energy and  phonon dispersion calculations, respectively. We applied the Gaussian smearing method with the smearing width of 0.05 Ryd. For the dynamical matrices of the $I4/mmm$ structure, we used a 2$\times$2$\times$2 {\bf q} grid, giving 8 wave vectors {\bf q} in the irreducible wedge of the first BZ. For the 2$H$- and 2$H'$-NbS$_2$, the {\bf q} grid meshes are 2$\times$4$\times$4, also giving 8 wave vectors.

Phonon dispersion curves (Figs.~\ref{fig:ph139}(b) and \ref{fig:ph194}(a)) do not show any imaginary frequencies, indicating dynamical stability of $I4/mmm$- and 2$H'$-NbS$_2$. So we believe $I4/mmm$- and 2$H'$-NbS$_2$ are both mechanically and dynamical stable. The phonon dispersion curve of 2$H$-NbS$_2$ are also presented in Fig.~\ref{fig:ph194}(b), compared to the experimental data.~\cite{Leroux2012} The agreement of our calculated phonon frequencies and the 300 K experimental data is quite good. By comparing the phonon dispersion curves of 2$H$- and 2$H'$-NbS$_2$, we note the phonon frequencies of 2$H'$-NbS$_2$ exhibit softening near to A point (along the $\Gamma$-A, A-L and H-A directions), implying its metastability compared to 2$H$ structure. This is consistent with the conclusions from the elastic constants calculations. 

\subsection{Electron-phonon coupling and superconductivity}
We calculated the superconducting transition temperature $T_c$ of NbS$_2$ using the Allen-Dynes~\cite{Allen1975} form of the McMillan~\cite{McMillan1968} equation,
\begin{equation}
  T_c=\frac{\omega_{\mathrm{ln}}}{1.2}\mathrm{exp}\left[-\frac{1.04(1+\lambda)}{\lambda-\mu^*(1+ 0.62\lambda)}\right],
  \label{mcmillan}
\end{equation}
where $\lambda$ ($=2\int^{\infty}_0\alpha^2F(\omega)/{\omega}\mathrm{d}\omega$) is the electron-phonon coupling constant, $\omega_{\mathrm{ln}}$ the logarithmic average frequency, and $\mu^*$  the Coulomb pseudopotential. The logarithmic average frequency is calculated by
\begin{equation}
\omega_{\mathrm{ln}}=\mathrm{exp}\{\frac{2}{\lambda} \int^{\infty}_0d\omega                      \alpha^2F(\omega)ln\omega/{\omega}\}.
\label{wln}
\end{equation}
The Eliashberg spectral function, $\alpha^2F(\omega)$, which measures the contribution of the phonons with frequency $\omega$ to the scattering of electrons,~\cite{Eliashberg1962} can be     written as,~\cite{McMillan1968}
\begin{equation}
  \alpha^2F(\omega)=\frac{1}{2\pi N(\epsilon_F)} \sum_{q\nu} \frac{\gamma_{q\nu}}{\omega_{q\nu}} \delta(\omega-\omega_{q\nu}),
  \label{a2F}
\end{equation}
where $N(\epsilon_F)$ is the EDOS at the Fermi level. The linewidth of the phonon mode was       calculated from,~\cite{McMillan1968}
\begin{equation}
  \gamma_{q\nu}=2\pi \omega_{q\nu}\sum_{kjj'}|g^{q\nu}_{k+qj',kj}|^2\delta(\epsilon_{kj}-        \epsilon_F)\delta(\epsilon_{k+qj'}-\epsilon_F),
  \label{}
\end{equation}
where $g^{q\nu}_{k+qj',kj}$ is the electron-phonon coupling matrix element. 
The Coulomb pseudopotential $\mu^*$ was taken the typical value 0.10 in all the superconducting critical temperatures ($T_c$) calculations. 

The calculated $T_c$ of 2$H$- and $I4/mmm$-NbS$_2$ are plotted in Fig.~\ref{fig:tc}, compared with recent experimental data.~\cite{Tissen2013} The resulting $T_c$s of 2$H$-NbS$_2$ are in very good agreement with experiment and increase with pressure. It is interesting that the $T_c$ of $I4/mmm$ structure is higher than that of 2$H$ structure and decreases with pressure. This is resulted from the stronger electron-phonon coupling coefficients $\lambda$ in $I4/mmm$-NbS$_2$ (Fig.~\ref{fig:tc}). The phonon calculations indicate that $I4/mmm$ is unstable below 10 GPa. The highest $T_c$ of $I4/mmm$ (at 10 GPa) is 17.83 K. From the electronic energy band structure of $I4/mmm$-NbS$_2$ (Figs.~\ref{fig:ph139}(a)), we note it is metallic. It is previously known that 2$H$-NbS$_2$ is also metallic, so pressure does not change the metallic properties of NbS$_2$, but enhances the electron-phonon coupling effects and thus increases the the superconducting critical temperature.

\section{Conclusions}
\label{concl}
In conclusion, we predicted three new 2$H'$-, $P3m1$-, and $I4/mmm$-NbS$_2$ structures using the MAC crystal structure prediction technique. The new 2$H'$-NbS$_2$ can be formed by shifting the layer of atoms along typical direction parallel to the layer plane. Based on enthalpy calculations, we found 2$H$-NbS$_2$ transits to the tetragonal $I4/mmm$ structure at 26 GPa. The new bondings in $I4/mmm$ form a [NbS$_8$] hexahedron, which has not been reported in TMD crystals. More interestingly, the superconducting temperature of $I4/mmm$-NbS$_2$ is higher than that of 2$H$-NbS$_2$ and decreases as pressure increases, resulted from the stronger electron-phonon coupling coefficients $\lambda$ in $I4/mmm$-NbS$_2$. In the stability region of $I4/mmm$ structure, the highest $T_c$ is 17.83 K.


\section{acknowledgments}
The research was supported by the National Natural Science Foundation of China (11104127, 11104227), the NSAF of China under grant
No. U1230201/A06, the Project 2010A0101001 funded by CAEP, and the Science Research Scheme of Henan Education Department under Grand No. 2011A140019.

\newpage
\bibliographystyle{apsrev4-1}
%
\end{document}